\documentclass[onecolumn,superscriptaddress,aps,pra,10pt,notitlepage]{revtex4-1}
\usepackage{color}
\usepackage{graphicx}
\usepackage{subfigure}
\usepackage{amsmath,amssymb}
\usepackage{ragged2e}
\usepackage{keyval}
\usepackage{braket}
\usepackage{natbib}
\usepackage{textcomp}
\usepackage[colorlinks,allcolors=blue]{hyperref}
\usepackage{bm}
\usepackage{todonotes}
\usepackage{array}
\newcolumntype{x}[1]{>{\centering\hspace{0pt}}p{#1}}
\newcommand{\tn}{\tabularnewline}

\newcommand{\ph}{\ensuremath{\varphi}}
\newcommand{\ii}{\ensuremath{\mathrm{i}}}
\newcommand{\ee}{\ensuremath{\mathrm{e}}}
\newcommand{\lt}{\ensuremath{\tilde{l}}}
\newcommand{\pt}{\ensuremath{\tilde{p}}}

\newcommand{\abs}[1]{\ensuremath{\left| #1 \right|}}

\begin{document}

\title{Recovering Quantum Information in Orbital Angular Momentum of Photons by Adaptive Optics}
\author{Jose Raul \surname{Gonzalez Alonso}}
\email[Electronic address: ]{jrgonzal@usc.edu}
\author{Todd A. Brun}
\email[Electronic address: ]{tbrun@usc.edu}
\affiliation{Department of Physics
and Astronomy, University of Southern California, Los Angeles, California
90089-0484, USA} 
\begin{abstract}
Orbital angular momentum of photons is an intriguing system for the storage and transmission of quantum information, but it is rapidly degraded by atmospheric turbulence.  We explore the ability of adaptive optics to compensate for this disturbance by measuring and correcting cumulative phase shifts in the wavefront.  These shifts can be represented as a sum of Zernike functions; we analyze the residual errors after correcting up to a certain number of Zernike modes when an orbital angular momentum state is transmitted through a turbulent atmosphere whose density fluctuations have a Kolmogorov spectrum. We approximate the superoperator map that represents these residual errors and find the solution in closed form. We illustrate with numerical examples how this perturbation depends on the the number of Zernike modes corrected and the orbital angular momentum state of the light.
\end{abstract}
\maketitle

\section{Introduction}\label{sec:intro}
Photons are very interesting carriers for quantum information since they are relatively easy to produce and transmit.  Most commonly, the polarization of photons---a two dimensional space---is the choice for encoding qubits in free space, though dual-rail and time-bin encodings are also common (particularly for transmission through optical fiber). However, higher dimensional spaces can be obtained with the use of the orbital angular momentum (OAM) of photons \cite{Allen-Orbital-1992-0,Yao-Orbital-2011-0, Willner-Optical-2015-0}, which could potentially allow larger channel capacities and increased key generation rates for protocols such as quantum key distribution.

Unfortunately, OAM of photons is highly sensitive to atmospheric turbulence \cite{Paterson-Atmospheric-2005-0,Gopaul-The-effect-2007-0,Roux-Decoherence-2010-0,Anguita-Turbulence-induced-2008-0,Tyler-Influence-2009-0,Sheng-Effects-2012-0}, and a method to protect quantum information encoded in the OAM of photons while they travel in a turbulent atmosphere remains elusive \cite{Gonzalez-Alonso-Protecting-2013-0}. If OAM photons are ever to be used in quantum communications, it is imperative to properly understand the processes they undergo in a turbulent atmosphere  \cite{Fried-Statistics-1965-0,Fried-Optical-1966-0,Wheelon-Electromagnetic-2001-0,Wheelon-Electromagnetic-2003-0,Gibson-Free-space-2004-0,Andrews-Laser-2005-0,Forbes-Laser-2014-0,Paterson-Atmospheric-2005-0,Gopaul-The-effect-2007-0,Anguita-Turbulence-induced-2008-0,Gbur-Vortex-2008-0,Tyler-Influence-2009-0,Roux-Decoherence-2010-0,Pors-Transport-2011-0,Sheng-Effects-2012-0,Ibrahim-Parameter-2014-0,Leonhard-Universal-2014-0,Jimenez-Farias-Resilience-2015-0,Roux-Non-Markovian-2015-0,Krenn-Twisted-2016-0} and how to compensate for them. 

One possibility to mitigate the effects of turbulence is to use adaptive optics \cite{Noll-Zernike-1976-0,Dai-Zernike-2007-0,Tyson-Principles-2010-0}. While this possiblity has been explored for classical communication \cite{Ren-Adaptive-2014-0} it still remains to be studied thoroughly in the quantum case.  In this paper, we explore the effects of atmospheric turbulence with adaptive optics on both the radial and azimuthal degrees of freedom \cite{Plick-The-Forgotten-2013-0,Chen-Characterizing-2016-0} of OAM photons.  For this purpose, we model atmospheric turbulence using the Kolmogorov spectrum, and the correction effects using Zernike functions \cite{Noll-Zernike-1976-0}.

In Sec.~\ref{sec:zernike} we expand the effects of turbulence in Zernike functions, and we use them in
Sec.~\ref{sec:calc_matrix_elems} to calculate the residual errors after the adaptive optics correction in a first order expansion.  We describe these errors as a superoperator acting on the state of the input photons, and derive integral expressions for the matrix elements of this superoperator.  In Sec.~\ref{sec:solving_integrals} we show how to calculate these integrals, and in Sec.~\ref{sec:num_ex} we illustrate our discussion with numerical examples.

\section{Atmospheric Turbulence, Adaptive Optics, and Zernike Functions}\label{sec:zernike}

As a photon propagates through a turbulent atmosphere, it encounters small fluctuations in the density (and hence the index of 
refraction) of the air.  These fluctuations vary across the wavefront of the photons.  We model the cumulative effects of turbulence by a 
spatially-varying phase change $\ph(r, \theta)$, where $r$ and $\theta$ are cylindrical coordinates across the wavefront.  (A third 
coordinate, $z$, represents the distance along the beam, which is generally the distance from the transmitter to the receiver.) The 
effects of turbulence can be represented semiclasically by a superator $\hat{\mathrm{T}}_{\ph}$ such that if
$\braket{\mathbf{r}|\psi}$ represents the wavefunction of a state with OAM, then the wavefunction of the state after one realization
of the noise is
\cite{Paterson-Atmospheric-2005-0,Gonzalez-Alonso-Protecting-2013-0}
\begin{align}
\Braket{\mathbf{r}|\hat{\mathrm{T}}_{\ph}|\psi} = \exp\left(\ii \ph (r,\theta)\right)\Braket{\mathbf{r}|\psi}.
\end{align}
The methods of adaptive optics estimate this phase $\ph(r,\theta)$ and then compensate for its effects.

These methods were originally developed for astronomy \cite{Tyson-Principles-2010-0}, but have more recently been applied to classical communication by OAM of photons through free space \cite{Ren-Adaptive-2014-0}.   Adaptive optics works by sending a bright pulse of light in a standard state through the same volume of air immediately before the communication pulse.  Because air moves slowly compared to light, the fluctuations encountered by this probe beam will be very close to that encountered by the communication pulse.  By measuring how the probe beam is distorted, the phase function $\ph(r, \theta)$ can estimated, and active optical elements can apply a compensating phase shift to cancel out the distortion.

How is this phase estimated?  One approach is to expand $\ph(r, \theta)$  in terms of an orthogonal set of functions that are defined in the receiving aperture of a system. Such a set is given by the Zernike functions $\{Z_k(r/R,\theta)\}$ \cite{Born-Principles-1999-0}, which are defined on a disk of radius $R$.  In terms of these functions we write the phase as
\begin{equation}
\ph(r, \theta) =\sum_{k=1}^{\infty} a_k Z_k\left(\frac{r}{R},\theta\right) .
\label{eq:zernike_phase}
\end{equation}
How do we define the functions $\{Z_k(r/R,\theta)\}$?  We assume $0\le r \le R$.  For integers $n,m$ such that $n\ge 0$, $n\ge \abs{m}$ and $n-\abs{m}$ is even, we can define a joint index $k$ using the ordering conventions in \cite{Noll-Zernike-1976-0}.   For $m\ne 0$ and $k$ even:
\begin{equation}
Z_k\left(\frac{r}{R},\theta\right) = \frac{1}{R}\sqrt{\frac{2(n+1)}{\pi}} P^{\abs{m}}_{n} \left(\frac{r}{R}\right) \cos(m \ph) .
\label{eq:zernike_pol_m_pos}
\end{equation}
For $m\ne0$ and $k$ odd:
\begin{equation}
Z_k\left(\frac{r}{R},\theta\right) = \frac{-1}{R}\sqrt{\frac{2(n+1)}{\pi}} P^{\abs{m}}_{n} \left(\frac{r}{R}\right) \sin(m \ph) .
\label{eq:zernike_pol_m_neg}
\end{equation}
Finally, for $m=0$:
\begin{equation}
Z_k\left(\frac{r}{R},\theta\right) = \frac{1}{R}\sqrt{\frac{n+1}{\pi}} P^{0}_{n} \left(\frac{r}{R}\right).
\label{eq:zernike_pol_m_zero}
\end{equation}
The polynomials $P^{\abs{m}}_n$ are given by
\begin{align}
P^{\abs{m}}_n \left(\frac{r}{R}\right) &= \sum_{s=0}^{\frac{n-\abs{m}}{2}} (-1)^s \binom{n-s}{s}\binom{n-2s}{\frac{n-\abs{m}}{2}-s} \left(\frac{r}{R}\right)^{n-2s}.
\label{eq:zernike_rad_pol}
\end{align}

The relationship between the integer parameters $m$ and $n$ and the joint index $k$ is somewhat complicated.  Defining
\[
T = \frac{n(n+1)}{2} , \ \ \ 
T_2 = T \mod 2 ,\ \ \ 
m_0 = \abs{m} \mod 2, \ \ \ 
m_1 = \abs{m} - 1 \mod 2 ,
\]
the joint index $k$ as used in \cite{Noll-Zernike-1976-0} is given by
\begin{widetext}
\begin{align}
k = 1 + T + \abs{m} - (1 - 
     \delta_{\abs{m},0})
     \left(
     H(m)\left[T_2 m_0 + (1 - T_2)m_1\right]
     + H(-m)\left[T_2 m_1 + (1-T_2) m_0\right]
     \right),
\end{align}
\end{widetext}
where
\begin{align}
H(x) =
\begin{cases}
0, &\text{if } x< 0 ,\\
1, &\text{if } x\geq 0 .
\end{cases}
\end{align}
If $m>0$, then $k$ is even, and if $m<0$, then $k$ is odd. In the case where $m=0$, the parity of $k$ can be either even or odd.  The index $k$ is constructed in such a way that, for a given $n$, indices with a smaller value of \abs{m} are smaller. In what follows, to maintain clarity in switching back and forth between the integer parameters $m$ and $n$ and the joint index $k$, we will denote by $n_k$ and $m_k$ the two integers corresponding to the particular joint index $k$.

With the definitions \eqref{eq:zernike_pol_m_pos}, \eqref{eq:zernike_pol_m_neg}, and \eqref{eq:zernike_pol_m_zero}, the Zernike functions form an orthonormal set on the disk of radius $R$:
\begin{align}
\int_{0}^{R}\int_{-\pi}^{\pi} r\mathrm{d}r\, \mathrm{d}\theta\, Z_k\left(\frac{r}{R},\theta\right)
Z_{\tilde{k}}\left(\frac{r}{R},\theta\right) = \delta_{k,\tilde{k}} .
\end{align}
Since the Zernike functions form a complete orthonormal set on the disk of radius $R$, we can always expand the phase change $\ph(r, \theta)$ using them, as in Eq.~\eqref{eq:zernike_phase}.  The main task in adaptive optics is then to estimate the coefficients of the expansion $a_k$ up to a certain number of Zernike modes $J$, and use this information to eliminate the aberrations due to the first $J$ modes.  In other words, an experimental procedure yields a correction phase $\ph_c$ given by
\begin{align}\label{eq:correction_AO}
\varphi_c (r,\theta) = \sum_{k=1}^{J} a_k Z_k\left(\frac{r}{R},\theta\right)
\end{align}
which is then subtracted from $\ph(r,\theta)$, leaving a residual phase \cite{Noll-Zernike-1976-0}:
\begin{align}\label{eq:residual_AO}
\begin{split}
\ph_A (r,\theta) &= \ph(r,\theta) - \ph_c (r,\theta)\\
&= \sum_{k=J+1}^{\infty} a_k Z_k\left(\frac{r}{R},\theta\right) .
\end{split}
\end{align}
Generally, the coefficients $a_k$ are estimated using a procedure such as the Shack-Hartmann wavefront sensing technique \cite{Roggemann-Imaging-1996-0,Tyson-Principles-2010-0}. However, because of the helical nature of the wavefronts of OAM states, this must be modified to incorporate a bright probe beam in order to do the wavefront estimation \cite{Ren-Adaptive-2014-0} required in the adaptive optics correction procedure.  The correction itself is done using fast active optics.

\section{Calculating the Matrix Elements of the Superoperator Representation of AO and Turbulence} \label{sec:calc_matrix_elems}

We are interested in protecting the quantum information initially encoded in an eigenstate of OAM from the effects of turbulence with the help of adaptive optics. We write the input state in terms of basis vectors $\ket{l_0,p_0}$:
\begin{align}\label{eq:wavefnOAM}
        \Braket{\mathbf{r}|l_0,p_0} = \frac{1}{\sqrt{2 \pi}}R_{l_0,p_0}(r,z)\exp(\ii l_0\theta),
\end{align}
where $r^2 = x^2 + y^2$, $\theta = \arctan\left(\frac{y}{x}\right)$. We assume the state propagates in the $z$ direction.  In what follows, we will use  the Laguerre-Gauss functions $R_{l_0,p_0}(r,z)$ \cite{Allen-Orbital-1992-0}:
\begin{equation}
R_{l_0,p_0}(r,z) = \frac{2 A_{l,p}}{w(z)}\left(\frac{\sqrt{2} r}{w(z)}\right)^{\abs{l_0}} L_{p_0}^{\abs{l_0}}\left(\frac{2r^2}{w(z)^2}\right)\ee^{-r^2/w(z)^2}
\ee^{-\ii kr^2/[2R(z)]}\ee^{ \ii(2p_0+\abs{l_0} + 1)\arctan(z/z_R)},
\label{eq:radialpartOAM}
\end{equation}
where $w(z)=w_0\sqrt{1+(z/z_R)^2}$ is the beam width, $R(z)=z[1+(z_R/z)^2]$ is the radius of wave-front curvature, and $z_R = \frac{1}{2}kw_0^2$ is the Rayleigh range. The quantity $\arctan(z/z_R)$ is known as the Gouy phase, and the normalization constant $A_{l,p}$ is
\begin{align}
A_{l,p} = \sqrt{\frac{p!}{(p+\abs{l})!}} .
\label{eq:LG_norm_const}
\end{align}
The functions $L_{p}^{\abs{l}}(x)$ are generalized Laguerre polynomials:
\begin{align}
 L_p^{\abs{l}} (x) = \sum_{i=0}^p (-1)^i \binom{p+\abs{l}}{p-i} \frac{x^i}{i!} .
 \label{eq:L_gen_pols}
\end{align}
Similarly to our earlier work in \cite{Gonzalez-Alonso-Protecting-2013-0}, the combined effects of turbulence and adaptive optics corrections on a basis state \eqref{eq:wavefnOAM} are represented by an operator $\hat{\mathrm{A}}_{\ph_A}$ such that
    \begin{equation}\label{eq:turAObeffect}
        \braket{\mathbf{r} | \hat{\mathrm{A}}_{\ph_A} | l_0,p_0} = \exp\left(\ii \ph_A (r,\theta)\right)\Braket{\mathbf{r}|l_0,p_0}.
    \end{equation}
However, this only describes the change of state for a particular realization of the noise.  We must make an ensemble average over $\ph_A(r,\theta)$ to find the superoperator representing the residual error process after the adaptive optics.

The OAM eigenstates in Eq.~\eqref{eq:wavefnOAM} form a complete basis. Therefore, we can use them to expand the state after the effects of the turbulent atmosphere and adaptive optics.  In general, after averaging over the noise the state will be mixed, so we represent it as a density matrix
\[
\rho = \sum_{l,l',p,p'} \rho_{l,p; l',p'} \ket{l,p}\bra{l',p'} .
\]
We will represent the effects of turbulence and adaptive optics after averaging using an operator, which we will denote by $\mathcal{\hat{A}}_{\ph_A}$, such that its matrix representation $\mathbf{A}$ satisfies
\begin{align}\label{eqn:matrix_rep_A}
 \left(\hat{\mathcal{A}}_{\ph_A} \rho \right)_{(\lt,\pt,\lt',\pt')} = \sum_{l,p,l',p'} \mathbf{A}_{(\lt,\pt,\lt',\pt'),(l,p,l',p')} \rho_{(l,p,l',p')}.
 \end{align}
In what follows, we describe how to obtain the matrix elements of $\mathbf{A}$ by explicitly doing the averaging process and calculating the reqquired integrals. By linearity, it suffices to know how the noise superoperator acts on outer products of the form $\Ket{l,p}\Bra{l',p'}$:
\begin{align}\label{eq:genturbAOeffect}
\begin{split}
\Ket{l,p}\Bra{l',p'} \mapsto & \frac{1}{4\pi^2} \sum_{\lt,\pt,\lt',\pt'} \iiiint r \, dr\, d\theta\, r'\, dr'\, d\theta'
  \overline{R_{\lt,\pt}}(r,z) R_{l,p}(r,z) \exp\left[\ii\left[ \theta \left( l-\lt \right) + \ph_A(r,\theta) \right]\right]\\
&\times R_{\lt',\pt'}(r',z) \overline{R_{l',p}}(r',z) \exp\left[-\ii\left[ \theta' \left( l'-\lt' \right) + \ph_A(r',\theta') \right]\right] \Ket{\lt,\pt}\Bra{\lt',\pt'}.
\end{split}
\end{align}
Since the atmospheric variations in the refraction index are random, we take the ensemble average $\mathbb{E}[\cdot]$ of \eqref{eq:genturbAOeffect}. Because of the linearity of $\mathbb{E}[\cdot]$, we need an expression for $\mathbb{E}\left[ \exp\{\ii \left(\ph_A(r,\theta) - \ph_A(r',\theta')\right)\} \right]$ for us to obtain the ensemble average.  Assuming that $\ph_A$ is a Gaussian random variable with zero mean, it is straightforward to show that the ensemble average simplifies to 
\begin{equation}
\label{eq:ensavgrandfluctAO}
\mathbb{E}\left[ \exp\{\ii \left(\ph_A(r,\theta) - \ph_A(r',\theta')\right)\} \right]
= \exp\left\{ -\frac{1}{2} \mathbb{E}\left[\left(\ph_A(r,\theta) - \ph_A(r',\theta')\right)^2\right]\right\}.
\end{equation}
If we suppose that the residual effects of the uncorrected noise are small, then we consider only terms up to first order in the series expansion of the exponential:
\begin{equation}
\exp\left\{ -\frac{1}{2} \mathbb{E}\left[\left(\ph_A(r,\theta) - \ph_A(r',\theta')\right)^2\right]\right\}
\approx 1 - \frac{1}{2}\mathbb{E}\left[\left(\ph_A(r,\theta) - \ph_A(r',\theta')\right)^2\right].
\end{equation}
Expanding $\ph_A$ in terms of Zernike functions as in Eq. \eqref{eq:residual_AO}, we can write
\begin{widetext}
\begin{equation}
\mathbb{E}\left[\left(\ph_A(r,\theta) - \ph_A(r',\theta')\right)^2\right] = 
\sum_{k,\tilde{k}=J+1}^{\infty} \mathbb{E}\left[a_k a_{\tilde{k}}\right]
\left(Z_k \left(\frac{r}{R},\theta\right) - Z_k \left(\frac{r'}{R},\theta'\right)\right)
\left(Z_{\tilde{k}} \left(\frac{r}{R},\theta\right) - Z_{\tilde{k}} \left(\frac{r'}{R},\theta'\right)\right) .
\end{equation}
\end{widetext}

The covariance of the expansion coefficients can be calculated explicitly using the Fourier transforms of the Zernike functions. If $n_k+n_{\tilde{k}}\ge 2$, then $\mathbb{E}\left[a_k a_{\tilde{k}}\right]$ is given by
\begin{equation}
\mathbb{E}\left[a_k a_{\tilde{k}}\right] = (-1)^{\frac{1}{2}(n_{\tilde{k}} - n_k)} M R^2
\left(\frac{R}{r_0}\right)^{\frac{5}{3}}\sqrt{(n_k+1)(n_{\tilde{k}}+1)}
\delta_{m_k,m_{\tilde{k}}} I_{n_k,n_{\tilde{k}}} ,
\label{eq:cov_expcoefsAO}
\end{equation}
where
\begin{equation}
M = \frac{4\sqrt{2}\left(\frac{3}{5}\Gamma\left(\frac{6}{5}\right)\right)^{5/6}\Gamma\left(\frac{11}{6}\right)^2}{\pi^{11/3}} \approx 0.04579117421711036 ,
\end{equation}
and 
\begin{equation}
r_0 = \left(\frac{16.6}{\lambda^2}\int_{L} \mathrm{d}\ell\, C_n^2\right)^{-3/5}
\label{eq:friedparameter}
\end{equation}
is called the Fried parameter, which has dimensions of length \cite{Fried-Statistics-1965-0}.  (Because of the convention we used in Eqs.~(\ref{eq:zernike_pol_m_pos}--\ref{eq:zernike_pol_m_zero}), there is an extra factor of $\pi R^2$ compared to the expression in Ref. \cite{Noll-Zernike-1976-0}.) In \eqref{eq:friedparameter}, $\lambda$ is the wavelength, $L$ is the propagation path, $\ell$ is a length element along the propagation path, and $C_n^2$ is called the atmospheric  refractive index structure constant and has units of $L^{-\frac{2}{3}}$. In spite of being called a  constant, $C_n^2$ depends on altitude, pressure, temperature and may vary along the path \cite{Andrews-Laser-2005-0}. However, for the horizontal path free-space propagation, it may be approximated by a constant \cite{Andrews-Laser-2005-0}. Therefore, in what follows we approximate $r_0 \approx \left(16.6 C_n^2 z/\lambda^2\right)^{-3/5}$.
Finally, we will use that
\begin{widetext}
\begin{align}
I_{n_k,n_{\tilde{k}}} = 
\frac{\pi^{11/3}\Gamma\left(\frac{14}{3}\right)
\Gamma\left(\frac{1}{2}(n_k + n_{\tilde{k}} - \frac{14}{3}+3)\right)}
{2\Gamma\left(\frac{1}{2}(n_{\tilde{k}} -n_k + \frac{14}{3}+1)\right)
\Gamma\left(\frac{1}{2}(n_k - n_{\tilde{k}} + \frac{14}{3}+1)\right)
\Gamma\left(\frac{1}{2}(n_k+n_{\tilde{k}} + \frac{14}{3}+3)\right)
}.
\end{align}
\label{eq:int_res_exp_coefsAO}
\end{widetext}
It is perhaps worth noting that the covariance of the expansion coefficients in Eq.~\eqref{eq:cov_expcoefsAO} becomes small as $n_k+n_{\tilde{k}}$ becomes large. 

Putting all of the above together, and then taking the ensemble average, the expression in Eq.~\eqref{eq:genturbAOeffect} is approximately
\begin{widetext}
\begin{align}\label{eq:approxgenturbAOeffect}
\begin{split}
\Ket{l,p}\Bra{l',p'} \mapsto \Ket{l,p}\Bra{l',p'} 
- \frac{1}{8\pi^2} \sum_{\lt,\pt,\lt',\pt'} \sum_{k,\tilde{k}} & \iiiint r \, dr\, d\theta\, r'\, dr'\, d\theta'\,
\mathbb{E}\left[a_k a_{\tilde{k}}\right]
\exp\left[\ii\left[ \theta \left( l-\lt \right) - \theta' \left( l'-\lt' \right) \right]\right] \\
&\times \left(Z_k \left(\frac{r}{R},\theta\right) - 
Z_k \left(\frac{r'}{R},\theta'\right)\right)
\left(Z_{\tilde{k}} \left(\frac{r}{R},\theta\right) - 
Z_{\tilde{k}} \left(\frac{r'}{R},\theta'\right)\right) \\
&\times \overline{R_{\lt,\pt}}(r,z) R_{l,p}(r,z) R_{\lt',\pt'}(r',z) \overline{R_{l',p}}(r',z) \Ket{\lt,\pt}\Bra{\lt',\pt'}.
\end{split}
\end{align}
\end{widetext}
After some work we can obtain exact expressions for all the integrals in Eq.~\eqref{eq:approxgenturbAOeffect}.

\section{Solving the integrals}\label{sec:solving_integrals}

When doing the integration of the zeroth order term of the expansion (i.e. where no Zernike functions appear), we simply use the orthonormality of the Laguerre-Gauss functions to conclude that this term reduces to $\delta_{l,\tilde{l}}\delta_{p,\tilde{p}}\delta_{l',\tilde{l}'}\delta_{p',\tilde{p}'}$. Calculating the first order term (that is, the one with Zernike functions) of the expansion requires a bit more work. 

\subsection{First order term of the expansion: Angular Part of the Integration}

First, we note that since the covariance of the expansion coefficients includes $\delta_{m_k,m_{\tilde{k}}}$, then $k$ and $\tilde{k}$, as well as $n_k$ and $n_{\tilde{k}}$, have the same parity for non-zero terms. Hence, the angular part of the product of functions $Z_k$ and $Z_{\tilde{k}}$ will have the same functional form, and the integrals vanish unless $m_k = \tilde{m}_k$. Thus, there are only three possible cases for each of the angular forms of each product depending on whether $m_k$ is negative, positive, or zero.

All of these angular integrals take exactly the same form:
\[
F\left(l,l',\tilde{l},\tilde{l}',m_k\right) =
\int_{-\pi}^\pi \int_{-\pi}^{\pi} d\theta\,d\theta'\,\exp\left[\ii\left[ \theta \left( l-\lt \right) - \theta' \left( l'-\lt' \right) \right]\right] f(\theta) g(\theta') ,
\]
where the particular functions $f(\theta)$ and $g(\theta')$ depend on which product of Zernike functions is being integrated, and also on the value of $m_k$.  We have four possible products of $Z_k$ and $Z_{\tilde{k}}$ to analyze, which give us three distinct integrals; and for each case the result is different for $m_k > 0$, $m_k < 0$, and $m_k = 0$.  We summarize the results of all the angular integrals in the tables below.

\begin{table}[ht]
\caption{Angular integrals for $Z_{k} \left(\frac{r}{R},\theta\right)Z_{\tilde{k}} \left(\frac{r}{R},\theta\right)$}
\label{table:angular1}
\begin{center}
\begin{tabular}{|c|c|c|}
\hline
range of $m_k$ & $f(\theta)g(\theta')$ & result \\
\hline
$m_k < 0$ & $\sin(m_k\theta)^2$ & $F_1 \left(l,l',\tilde{l},\tilde{l}',m_k\right) = \begin{cases} 
    2\pi^2 & \text{if } l' = \tilde{l}' \text{ and } l = \tilde{l} \\
    -\pi^2 & \text{if } l' = \tilde{l}' \text{ and } l = \tilde{l}\pm 2 m_k \\
    0 & \text{otherwise}
\end{cases}$ \\
\hline
$m_k > 0$ & $\cos(m_k\theta)^2$ & $F_1 \left(l,l',\tilde{l},\tilde{l}',m_k\right) = \begin{cases} 
    2\pi^2 & \text{if } l' = \tilde{l}' \text{ and } l = \tilde{l} \\
    \pi^2 & \text{if } l' = \tilde{l}' \text{ and } l = \tilde{l}\pm 2 m_k \\
    0 & \text{otherwise}
\end{cases}$ \\
\hline
$m_k = 0$ & $1$ & $F_1 \left(l,l',\tilde{l},\tilde{l}',m_k\right) = \begin{cases} 
    4\pi^2 & \text{if } l' = \tilde{l}' \text{ and } l = \tilde{l} \\
    0 & \text{otherwise}
\end{cases}$ \\
\hline
\end{tabular}
\end{center}
\end{table}%

\begin{enumerate}
\item The first case is $Z_{k} \left(\frac{r}{R},\theta\right)Z_{\tilde{k}} \left(\frac{r}{R},\theta\right)$.  The results of the angular integrals are given in Table~\ref{table:angular1}.

\item The second case is $Z_{k} \left(\frac{r}{R},\theta\right)Z_{\tilde{k}} \left(\frac{r'}{R},\theta'\right)$.  The results of the angular integrals are given in Table~\ref{table:angular2}.

\item The third case is $Z_{k} \left(\frac{r'}{R},\theta'\right)Z_{\tilde{k}} \left(\frac{r}{R},\theta\right)$, whose integral is equal to that in case 2 because $m_k = m_{\tilde{k}}$.  The results of the angular integrals are also given in Table~\ref{table:angular2}.

\item The fourth case is $Z_{k} \left(\frac{r'}{R},\theta'\right)Z_{\tilde{k}} \left(\frac{r'}{R},\theta'\right)$.  
The results of the angular integrals are given in Table~\ref{table:angular3} and they are similar to those in Table
\ref{table:angular1} with the primed and unprimed variables and parameters switched.
\end{enumerate}

\begin{table}[ht]
\caption{Angular integrals for $Z_{k} \left(\frac{r}{R},\theta\right)Z_{\tilde{k}} \left(\frac{r'}{R},\theta'\right)$ and $Z_{k} \left(\frac{r'}{R},\theta'\right)Z_{\tilde{k}} \left(\frac{r}{R},\theta\right)$}
\label{table:angular2}
\begin{center}
\begin{tabular}{|c|c|c|}
\hline
range of $m_k$ & $f(\theta)g(\theta')$ & result \\
\hline
$m_k < 0$ & $\sin(m_k\theta) \sin(m_k\theta') $ & $F_{2,3} \left(l,l',\tilde{l},\tilde{l}',m_k\right) = \begin{cases} 
    \pi^2   & \text{if } l' = \tilde{l}' + m_k \text{ and } l = \tilde{l} + m_k \\
    \pi^2   & \text{if } l' = \tilde{l}' - m_k \text{ and } l = \tilde{l} - m_k \\
    -\pi^2  & \text{if } l' = \tilde{l}' + m_k \text{ and } l = \tilde{l} - m_k \\
    -\pi^2  & \text{if } l' = \tilde{l}' - m_k \text{ and } l = \tilde{l} + m_k\\
    0   & \text{otherwise}
\end{cases}$ \\
\hline
$m_k > 0$ & $\cos(m_k\theta) \cos(m_k\theta')$ & $F_{2,3} \left(l,l',\tilde{l},\tilde{l}',m_k\right) = \begin{cases} 
    \pi^2   & \text{if } l' = \tilde{l}' \pm m_k \text{ and } l = \tilde{l} \mp m_k \\
    0   & \text{otherwise}
\end{cases}$ \\
\hline
$m_k = 0$ & $1$ & $F_{2,3} \left(l,l',\tilde{l},\tilde{l}',m_k\right) = \begin{cases} 
    4\pi^2 & \text{if } l' = \tilde{l}' \text{ and } l = \tilde{l} \\
    0 & \text{otherwise}
\end{cases}$ \\
\hline
\end{tabular}
\end{center}
\end{table}%

\begin{table}[ht]
\caption{Angular integrals for $Z_{k} \left(\frac{r'}{R},\theta'\right)Z_{\tilde{k}} \left(\frac{r'}{R},\theta'\right)$}
\label{table:angular3}
\begin{center}
\begin{tabular}{|c|c|c|}
\hline
range of $m_k$ & $f(\theta)g(\theta')$ & result \\
\hline
$m_k < 0$ & $\sin(m_k\theta')^2$ & $F_4 \left(l,l',\tilde{l},\tilde{l}',m_k\right) = \begin{cases} 
    2\pi^2 & \text{if } l' = \tilde{l}' \text{ and } l = \tilde{l} \\
    -\pi^2 & \text{if } l' = \tilde{l}' \mp 2 m_k \text{ and } l = \tilde{l} \\
    0 & \text{otherwise}
\end{cases}$ \\
\hline
$m_k > 0$ & $\cos(m_k\theta')^2$ & $F_4 \left(l,l',\tilde{l},\tilde{l}',m_k\right) = \begin{cases} 
    2\pi^2 & \text{if } l' = \tilde{l}' \text{ and } l = \tilde{l} \\
    \pi^2 & \text{if } l' = \tilde{l}' \mp 2 m_k \text{ and } l = \tilde{l} \\
    0 & \text{otherwise}
\end{cases}$ \\
\hline
$m_k = 0$ & $1$ & $F_4 \left(l,l',\tilde{l},\tilde{l}',m_k\right) = \begin{cases} 
    4\pi^2 & \text{if } l' = \tilde{l}' \text{ and } l = \tilde{l} \\
    0 & \text{otherwise}
\end{cases}$ \\
\hline
\end{tabular}
\end{center}
\end{table}%

\subsection{First order term of the expansion: Radial Part of the Integration}

Now that we have completed the angular part of the integrals, we can proceed with the radial part. The different cases are the same as we used above. For simplicity, we will denote the products of the constants of normalization (and exponentials involving the Gouy phase) in Eqs.~(\ref{eq:zernike_pol_m_pos}--\ref{eq:zernike_pol_m_zero}) and \eqref{eq:radialpartOAM} as
\begin{align}\label{eq:all_norm_const}
\begin{split}
\mathcal{N}\left(l,p,l',p',\tilde{l},\tilde{p},\tilde{l}',\tilde{p}', k, \tilde{k}\right) =& \frac{16 \epsilon_{m_k}}{\pi R^2 w(z)^4} A_{l,p} A_{l',p'} A_{\tilde{l},\tilde{p}} A_{\tilde{l}',\tilde{p}'} \sqrt{\left(n_k + 1\right)\left(n_{\tilde{k}}+1\right)}\\
&\times \ee^{\left[\ii\arctan\left(\frac{z}{z_R}\right)\left(2p+\abs{l}-2p'-\abs{l'}-2\tilde{p}-\abs{\tilde{l}}+2\tilde{p}'+\abs{\tilde{l}'}\right)\right]}
\end{split}
\end{align}
where
\begin{align}
\epsilon_{m_k} =
\begin{cases}
    2, & \text{if } m_k \ne 0, \\
    1, & \text{if } m_k = 0.
\end{cases}
\end{align}

In the integrals that follow, we integrate the product of the respective Laguerre and Zernike functions term by term using the substitutions $u=\frac{2r^2}{w(z)^2}$ and $v=\frac{2{r'}^2}{w(z)^2}$. Additionally, when possible we have used the conditions that arise from the angular integrals (and the covariance of the expansion coefficients) relating the different indices (since the angular integrals vanish except for certain combinations of indices). To get an exact expression, we integrate the product of all the polynomials involved term by term using Eqs.~\eqref{eq:zernike_rad_pol}  and \eqref{eq:L_gen_pols}.

The integrals all take the following form:
\begin{widetext}
\begin{align}
\begin{split}
\mathcal{N} \int_{0}^{R}dr' \int_{0}^{R} dr\, & r r' 
L_{p}^{\abs{l}}\left(\frac{2r^2}{w(z)^2}\right) L_{\tilde{p}}^{\abs{\tilde{l}}}\left(\frac{2r^2}{w(z)^2}\right)
L_{p'}^{\abs{l'}}\left(\frac{2{r'}^2}{w(z)^2}\right)L_{\tilde{p}'}^{\abs{l'}}\left(\frac{2{r'}^2}{w(z)^2}\right)
f(r) g(r') \ee^{-\frac{2r^2}{w(z)^2}-\frac{2{r'}^2}{w(z)^2}} ,
\label{eq:radial_form}
\end{split}
\end{align}
\end{widetext}
where $f$ and $g$ are functions that will depend on which product of Zernike functions is being integrated.

\begin{enumerate}
\item The first case is $Z_{k} \left(\frac{r}{R},\theta\right)Z_{\tilde{k}} \left(\frac{r}{R},\theta\right)$.  All of the angular integrals of this case vanish except when $l'=\tilde{l}'$. The functions $f$ and $g$ in this case are
\begin{align}
f(r) &= \left(\frac{\sqrt{2}r}{w(z)}\right)^{|{l}|+|{\tilde{l}}|} P_{n_k}^{\abs{m_k}}\left(\frac{r}{R}\right) P_{n_{\tilde{k}}}^{\abs{m_k}}\left(\frac{r}{R}\right)\\
g(r') &= \left(\frac{\sqrt{2}r'}{w(z)}\right)^{2|{l'}|}
\end{align}
and the result of the integral is
\begin{equation}
\frac{\mathcal{N}w(z)^4}{16}\frac{\left(p' + \abs{l'}\right)!}{{p'}!}\delta_{p',\tilde{p}'} G_1 \left(l,p,\tilde{l},\tilde{p},k,\tilde{k}\right) ,
\end{equation}
where
\begin{widetext}
\begin{align}\label{eq:rad_part_int_1}
\begin{split}
&G_1 \left(l,p,\tilde{l},\tilde{p},k,\tilde{k}\right)=
\sum_{j=0}^{p}\sum_{\tilde{\jmath}=0}^{\tilde{p}}\sum_{j'=0}^{\frac{n_{k}-\abs{m_k}}{2}}
\sum_{\tilde{\jmath}'=0}^{\frac{n_{\tilde{k}}-\abs{m_k}}{2}}
\frac{(-1)^{\left(j+j'+\tilde{\jmath}+\tilde{\jmath}'\right)}}{j!\tilde{\jmath}!}\binom{p+\abs{l}}{p-j}
\binom{\tilde{p}+\abs{\tilde{l}}}{\tilde{p}-\tilde{\jmath}}\binom{n_k -j'}{j'}\binom{n_k - 2 j'}{\frac{n_k - \abs{m_k}}{2} - j'} \\
&\times \binom{n_{\tilde{k}}-\tilde{\jmath}'}{\tilde{\jmath}'}
\binom{n_{\tilde{k}}-2\tilde{\jmath}'}{\frac{n_{\tilde{k}}-\abs{m_k}}{2}-\tilde{\jmath}'}
\left(\frac{\sqrt{2}R}{w(z)}\right)^{\left(2j'+2\tilde{\jmath}'-n_k-n_{\tilde{k}}\right)}
\gamma\left( \frac{1}{2} \left( \abs{l} + \abs{\tilde{l}} + n_k  + n_{\tilde{k}} + 2 \left( j + \tilde{\jmath} - j' - \tilde{\jmath}' \right)\right)+1, \frac{2 R^2}{w(z)^2}\right).
\end{split}
\end{align}
\end{widetext}
In Eq. \eqref{eq:rad_part_int_1}, we have used the lower incomplete gamma function
\[
\gamma(\alpha,x) = \int_{0}^{x} \mathrm{d}x\, \ee^{-t} t ^{\alpha-1}
\]
to write the result of the term by term integration.

\item The second case is $Z_{k} \left(\frac{r}{R},\theta\right)Z_{\tilde{k}} \left(\frac{r'}{R},\theta'\right)$.  In this case, the combinations of indices where the different angular integrals vanish depend on the values of $m_k$, without a particular combination common to all values of $m_k$.  The functions $f$ and $g$ in this case are
\begin{align}
f(r) &= \left(\frac{\sqrt{2}r}{w(z)}\right)^{|{l}|+|{\tilde{l}}|} P_{n_k}^{\abs{m_k}}\left(\frac{r}{R}\right)\\
g(r') &= \left(\frac{\sqrt{2}r'}{w(z)}\right)^{|{l'}|+|{\tilde{l}'}|} P_{n_{\tilde{k}}}^{\abs{m_k}}\left(\frac{r'}{R}\right)
\end{align}
and after a tedious but straightforward calculation, we find the result of the integral:
\begin{equation}
\frac{\mathcal{N}w(z)^4}{16} G_2 \left(l,p,l',p',\tilde{l},\tilde{p},\tilde{l}',\tilde{p}',k,\tilde{k}\right)
\end{equation}
where
\begin{widetext}
\begin{align}\label{eq:rad_part_int_2}
\begin{split}
G_2 \left(l,p,l',p',\tilde{l},\tilde{p},\tilde{l}',\tilde{p}',k,\tilde{k}\right) = &
\left(\sum_{j=0}^{p} \sum_{\tilde{\jmath}=0}^{\tilde{p}}\sum_{j'=0}^{\frac{n_{k}-\abs{m_k}}{2}}
\frac{(-1)^{\left(j+\tilde{\jmath}+j'\right)}}{j!\tilde{\jmath}!}\binom{p+\abs{l}}{p-j}
\binom{\tilde{p}+\abs{\tilde{l}}}{\tilde{p}-\tilde{\jmath}}\binom{n_k -j'}{j'}
\binom{n_k - 2 j'}{\frac{n_k - \abs{m_k}}{2} - j'} \right.
\\
&\left.\times \left(\frac{\sqrt{2}R}{w(z)}\right)^{\left(2j'-n_k\right)}
\gamma\left( \frac{1}{2} \left( \abs{l} + \abs{\tilde{l}} + n_k + 2 \left( j + \tilde{\jmath} - j' \right)\right)+1, \frac{2 R^2}{w(z)^2}\right)\right)\\
&\times\left(
\sum_{i=0}^{p'}\sum_{\tilde{\imath}=0}^{\tilde{p}'}\sum_{i'=0}^{\frac{n_{\tilde{k}}-\abs{m_k}}{2}}
\frac{(-1)^{\left(i+\tilde{\imath}+i'\right)}}{i!\tilde{\imath}!}\binom{p'+\abs{l'}}{p'-i}
\binom{\tilde{p}'+\abs{\tilde{l}'}}{\tilde{p}'-\tilde{\imath}} 
\binom{n_{\tilde{k}}-i'}{i'}\binom{n_{\tilde{k}}-2i'}{\frac{n_{\tilde{k}}-\abs{m_k}}{2}-i'}
\right.\\
&\left.\times \left(\frac{\sqrt{2}R}{w(z)}\right)^{\left(2i'-n_{\tilde{k}}\right)}
\gamma\left( \frac{1}{2} \left( \abs{l'} + \abs{\tilde{l}'} + n_{\tilde{k}} + 2 \left( i + \tilde{\imath} - i' \right)\right)+1, \frac{2 R^2}{w(z)^2}\right)\right).
\end{split}
\end{align}
\end{widetext}
This solution also works for the $Z_{k} \left(\frac{r'}{R},\theta'\right)Z_{\tilde{k}} \left(\frac{r}{R},\theta\right)$ terms, by using \eqref{eq:rad_part_int_2} and switching $n_k\leftrightarrow n_{\tilde{k}}$.  In this case, we call the function $G_3$.

\item The third case is $Z_{\tilde{k}} \left(\frac{r'}{R},\theta'\right)Z_{\tilde{k}} \left(\frac{r'}{R},\theta'\right)$.  In this case, the angular part of the integration (calculated above) vanishes except when $l=\tilde{l}$, and the functions $f$ and $g$ are
\begin{align}
f(r) &= \left(\frac{\sqrt{2}r}{w(z)}\right)^{2|{l}|}\\
g(r') &= \left(\frac{\sqrt{2}r'}{w(z)}\right)^{|{l'}|+|{\tilde{l}'}|} P_{n_k}^{\abs{m_k}}\left(\frac{r'}{R}\right) P_{n_{\tilde{k}}}^{\abs{m_k}}\left(\frac{r'}{R}\right).
\end{align}

We can follow the same procedure we used to obtain \eqref{eq:rad_part_int_1} to find the result of the integral
\begin{equation}
\frac{\mathcal{N}w(z)^4}{16}\frac{\left(p + \abs{l}\right)!}{{p}!}\delta_{p,\tilde{p}} G_4 \left(l',p',\tilde{l}',\tilde{p}',k,\tilde{k}\right)
\end{equation}
where
\begin{widetext}
\begin{align}\label{eq:rad_part_int_4}
\begin{split}
&G_4 \left(l',p',\tilde{l}',\tilde{p}',k,\tilde{k}\right)=
\sum_{i=0}^{p'}
\sum_{\tilde{\imath}=0}^{\tilde{p}'}\sum_{i'=0}^{\frac{n_{k}-\abs{m_k}}{2}}\sum_{\tilde{\imath}'=0}^{\frac{n_{\tilde{k}}-\abs{m_k}}{2}}
\frac{(-1)^{\left(i+i'+\tilde{\imath}+\tilde{\imath}'\right)}}{i!\tilde{\imath}!}\binom{p'+\abs{l'}}{p'-i}
\binom{\tilde{p}'+\abs{\tilde{l}'}}{\tilde{p}'-\tilde{\imath}}\binom{n_k -i'}{i'}\binom{n_k - 2 i'}{\frac{n_k - \abs{m_k}}{2} - i'}\\
&\times \binom{n_{\tilde{k}}-\tilde{\imath}'}{\tilde{\imath}'}
\binom{n_{\tilde{k}}-2\tilde{\imath}'}{\frac{n_{\tilde{k}}-\abs{m_k}}{2}-\tilde{\imath}'}
\left(\frac{\sqrt{2}R}{w(z)}\right)^{\left(2i'+2\tilde{\imath}'-n_k-n_{\tilde{k}}\right)}
\gamma\left( \frac{1}{2} \left( \abs{l'} + \abs{\tilde{l}'} + n_k  + n_{\tilde{k}} + 2 \left( i + \tilde{\imath} - i' - \tilde{\imath}' \right)\right)+1, \frac{2 R^2}{w(z)^2}\right).
\end{split}
\end{align}
\end{widetext}
\end{enumerate}

\subsection{Putting it all together}

Now that we have obtained the results for all of the integrals, we can finally write the result of the first order expansion we did in \eqref{eq:approxgenturbAOeffect}. To this end, we use the following shorthand for the angular and radial results of the integration.  We take the definitions of $F_1$ from Table~\ref{table:angular1}, $F_2$ and $F_3$ from Table~\ref{table:angular2}, and $F_4$ from Table~\ref{table:angular3}.  For the radial part we use the definitions of $G_{1,2,3,4}$ from Eqs.~(\ref{eq:rad_part_int_1}--\ref{eq:rad_part_int_4}).  With all of this, we can write the combined effects of the turbulence and the adaptive optics correction as
\begin{widetext}
\begin{align}
&\Ket{l,p}\Bra{l',p'} \mapsto \Ket{l,p}\Bra{l',p'} - \frac{1}{8\pi^2}\sum_{\lt,\pt,\lt',\pt',k,\tilde{k}}
\mathcal{C}
\left(\frac{\delta_{p',\tilde{p}''}}{A_{l',p'}^2}F_1G_1 - F_2G_2 - F_3G_3 + 
\frac{\delta_{p,\tilde{p}}}{A_{l,p}^2}F_4G_4
\right)
\Ket{\lt,\pt}\Bra{\lt',\pt'}
\label{eq:approxgenturbAOeffectfinal}
\end{align}
\end{widetext}
where
\begin{align}\label{eq:all_norm_const_final}
\begin{split}
\mathcal{C} =
\mathbb{E}\left[a_k a_{\tilde{k}}\right] \mathcal{N} \frac{w(z)^4}{16}
=& \frac{\epsilon_{m_k}M}{\pi} A_{l,p} A_{l',p'} A_{\tilde{l},\tilde{p}} A_{\tilde{l}',\tilde{p}'}
\left(n_k + 1\right)\left(n_{\tilde{k}}+1\right) (-1)^{\frac{1}{2}(n_{\tilde{k}} - n_k)} \\
&\times I_{n_k,n_{\tilde{k}}}
\ee^{\left[\ii\arctan\left(\frac{z}{z_R}\right)\left(2p+\abs{l}-2p'-\abs{l'}-2\tilde{p}-\abs{\tilde{l}}+2\tilde{p}'+\abs{\tilde{l}'}\right)\right]}
\delta_{m_k,m_{\tilde{k}}}.
\end{split}
\end{align}
The indices $k,\tilde{k}$ are such that $k\ge J+1$, $\tilde{k}\ge J+1$ and $n_k + n_{\tilde{k}}\ge 2$. We use $J$ to denote the highest order of the Zernike functions used in the correction of the turbulence effects (see Eq.~\eqref{eq:residual_AO}). Additionally, all the OAM radial indices are non-negative integers, while the azimuthal indices can take positive or negative integer values.

\section{Numerical Examples}\label{sec:num_ex}
We can see from Eq. \eqref{eq:approxgenturbAOeffectfinal} that, up to first order, the map that represents the effects of the turbulence and the adaptive optics, which we have called $\mathcal{\hat{A}}_{\ph_A}$, is a perturbation to the identity map. How small this perturbation is, and how it behaves as we modify some of the parameters used in the description of both the turbulence and the Zernike expansion, is something we will study below using the same procedure we used previously in \cite{Gonzalez-Alonso-Protecting-2013-0}. Since the map is linear (a superoperator) and completely positive, we can represent it by a Choi matrix \cite{Choi-Completely-1975-0}. Diagonalizing the Choi matrix of this representation allows us to find a set of Kraus operators for the map.

We can characterize the perturbation by using the dimensionless parameters ${R}/{w(z)}$, ${w(z)}/{r_0}$, and ${z}/{z_R}$, and the number $J$ of modes used in the correction. Because we are doing a first order approximation to the map that represents the combined effects of the turbulence and the adaptive optics corrections, we will limit ourselves to values of the parameters that represent weak turbulence and small residual errors.

In what follows, we will illustrate numerically (using the library in \cite{Galassi-Gnu-Scientific-2003-0}) how changes in the parameters affect the map given by Eq. \eqref{eq:approxgenturbAOeffectfinal}.  To facilitate the analysis, we will restrict the sizes of the Hilbert spaces of the OAM states before and after the channel. Specifically, we will consider input states with azimuthal indices that satisfy $\abs{l_{in}}\le 3$ and radial indices such that $0 \le p_{in} \le 6$, and output states with azimuthal indices $\abs{l_{out}}\le 6$ and radial indices $0 \le p_{out} \le 6$.

To ensure the validity of the approximation used to derive Eq.~\eqref{eq:approxgenturbAOeffectfinal}, we have considered cases where the turbulence is weak by choosing a small value of $C_n^2$. Moreover, we also chose parameters in order to ensure that the scintillation (which we neglect) is weak. For the case of Gaussian beams, the scintillation is weak when the Rytov variance \cite{Andrews-Laser-2005-0}
\begin{align}\label{eq:rytov_var}
\sigma_R^2 = 1.637 t_z^\frac{5}{6}
\left(\frac{w(0)}{r_0}\right)^\frac{5}{3}
\end{align}
satisfies the condition \cite{Andrews-Laser-2005-0}
\begin{align}
\sigma_R^2 < \left(t_z + \frac{1}{t_z}\right)^\frac{5}{6},
\end{align}
where $t_z=\frac{z}{z_R}$.

For cases where no adaptive optics correction is applied (i.e. the number of modes corrected is zero), we have used the numerical procedure from our previous work \cite{Gonzalez-Alonso-Protecting-2013-0}. However, for all the cases in which there are correction being applied, we use Eq.~\eqref{eq:approxgenturbAOeffectfinal} to estimate the final state numerically with the help of the excellent GNU Scientific Library \cite{Galassi-Gnu-Scientific-2003-0}.

Just as in \cite{Gonzalez-Alonso-Protecting-2013-0}, we find that the spectrum of the Choi matrix of the map $\mathcal{\hat{A}}_{\ph_A}$ is dominated by its largest eigenvalue.  This eigenvalue corresponds to an error operator that is close to the identity.  The remaining nonzero eigenvalues mostly come in degenerate pairs. One of the error operators associated with these degenerate eigenvalues lowers the orbital angular momentum azimuthal number by a given amount, while the other operator raises it by the same amount. As the amount by which the OAM is raised or lowered increases, the magnitude of the eigenvalue decreases (making these errors weaker or less probable). These pairs of error operators can also raise or lower the radial number $p$.  The effects of these errors are more noticeable the larger the values of the azimuthal or radial numbers become. 

\begin{figure}
\subfigure[$\frac{R}{w(z)}=9.2088$, $\frac{w(z)}{r_0}=0.2165$, and $\frac{z}{z_R}=0.4234$]
{\includegraphics[width=0.45\columnwidth]{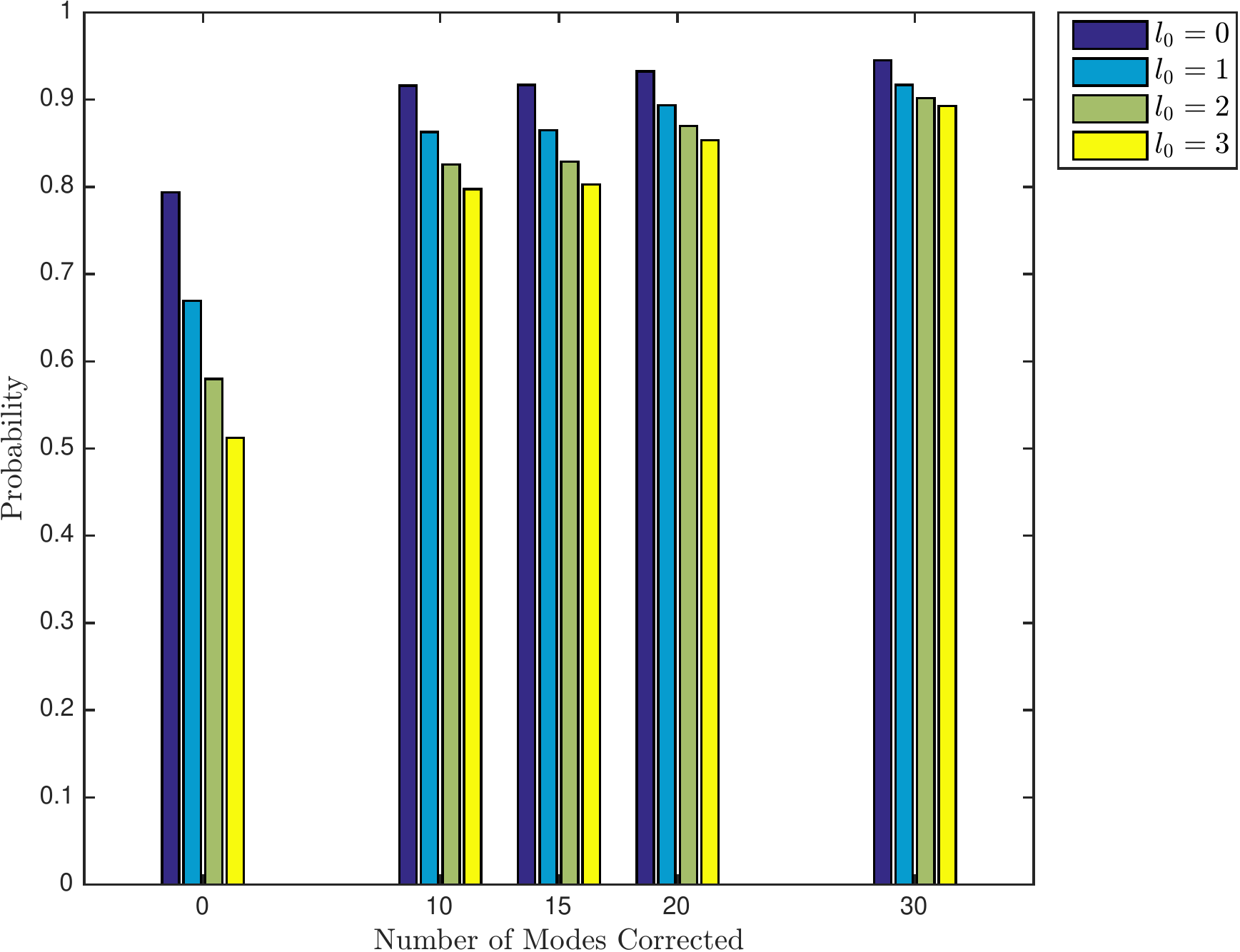}}
\subfigure[$\frac{R}{w(z)}=9.8596$, $\frac{w(z)}{r_0}=0.1167$, and $\frac{z}{z_R}=0.1693$]
{\includegraphics[width=0.45\columnwidth]{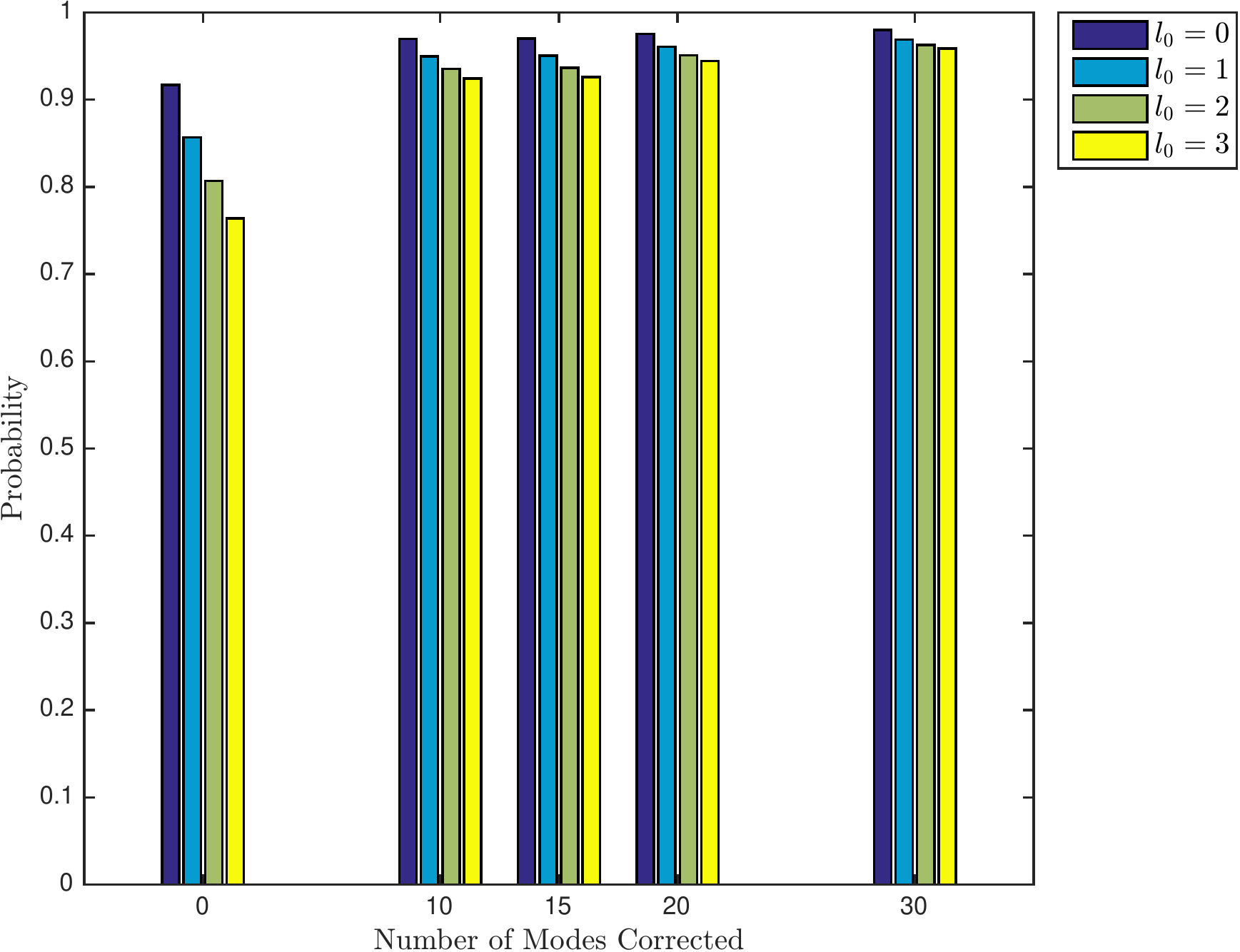}}
\caption{Probability to observe the initial state after turbulence and adaptive optics corrections.}
\label{fig:initial-l-effects}
\end{figure}

As an example of this dependence, consider a state that initially has $l=l'=l_0$ and $p=p'=0$.  In Fig.~\ref{fig:initial-l-effects} we see that as the value of $l_0$ increases, the probability that the output state is the same as the initial state decreases.  However, as the number of corrected modes increases, the probability that the input state is unchanged also increases, and the differences between these probabilities for different values of $l_0$ become smaller. (Unsurprisingly, this is also true as the strength of the turbulence decreases.)

\begin{figure}
\subfigure[$\frac{R}{w(z)}=9.2088$, $\frac{w(z)}{r_0}=0.2165$, and $\frac{z}{z_R}=0.4234$]
{\includegraphics[width=0.45\columnwidth]{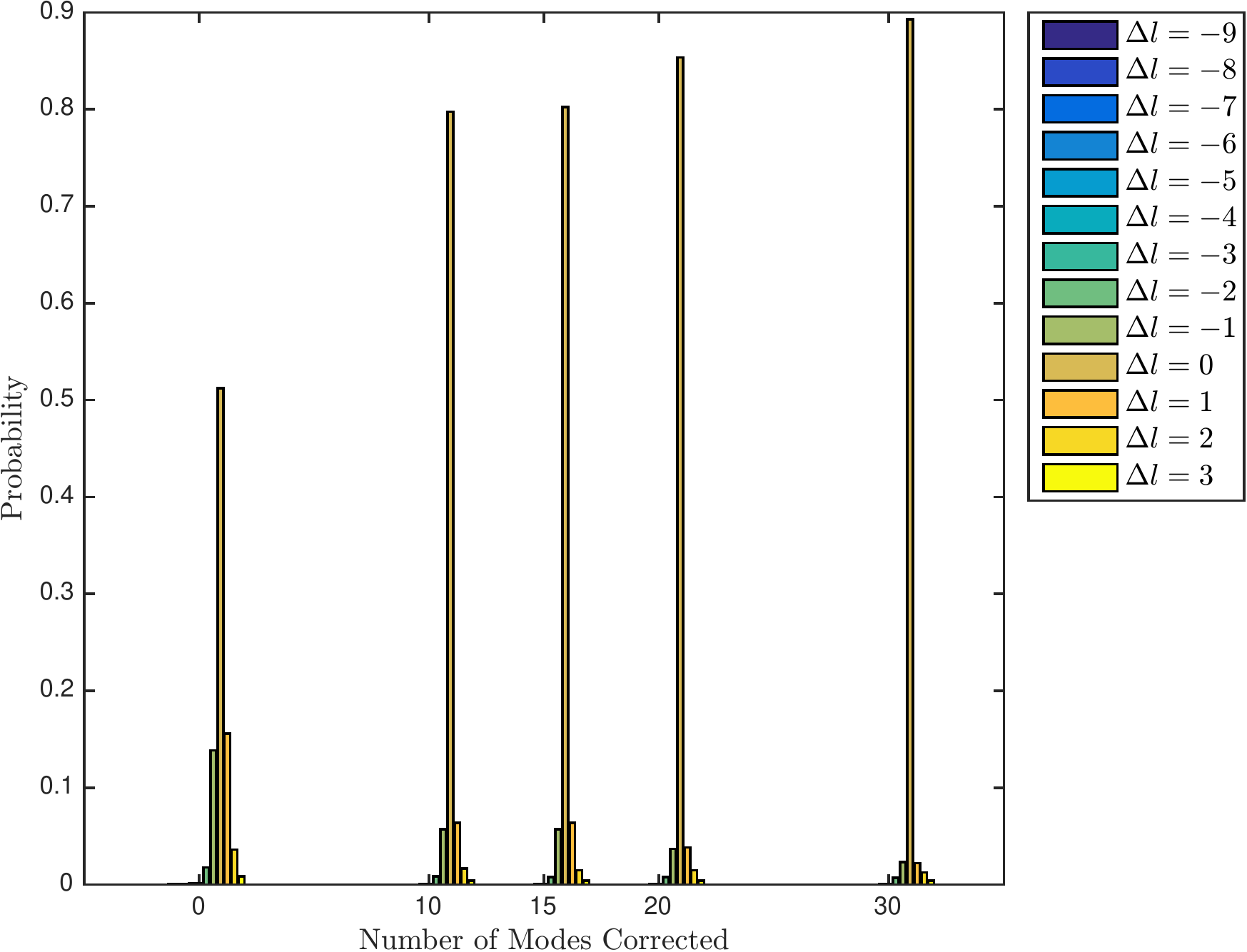}}
\subfigure[$\frac{R}{w(z)}=9.8596$, $\frac{w(z)}{r_0}=0.1167$, and $\frac{z}{z_R}=0.1693$]
{\includegraphics[width=0.45\columnwidth]{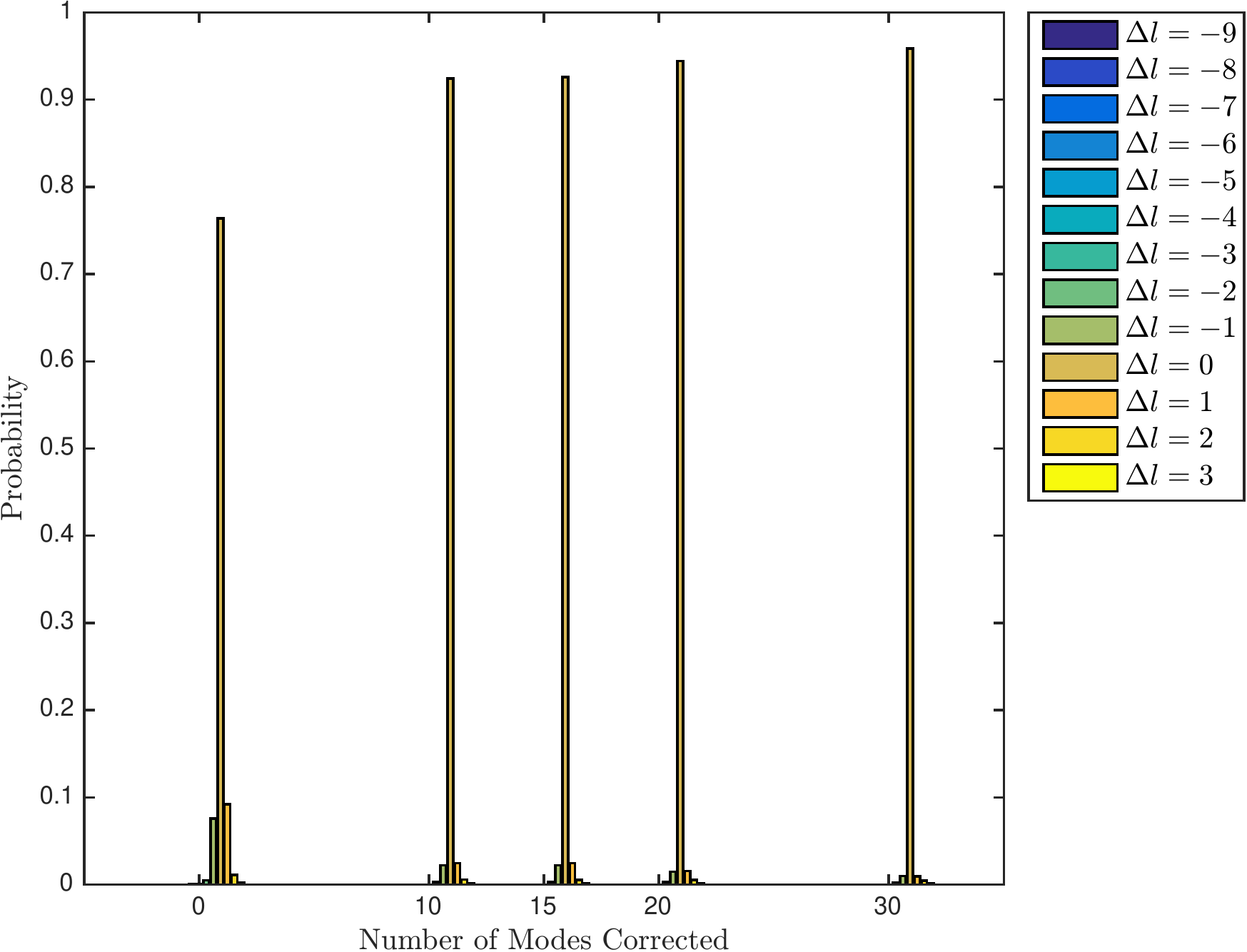}}
\caption{Probability to observe the state $\ket{3+\Delta l,0}$ after the turbulence and adaptive optics correction for initial state $\ket{3,0}$.}
\label{fig:l-spread}
\end{figure}

\begin{figure}
\subfigure[$\frac{R}{w(z)}=9.2088$, $\frac{w(z)}{r_0}=0.2165$, and $\frac{z}{z_R}=0.4234$]
{\includegraphics[width=0.45\columnwidth]{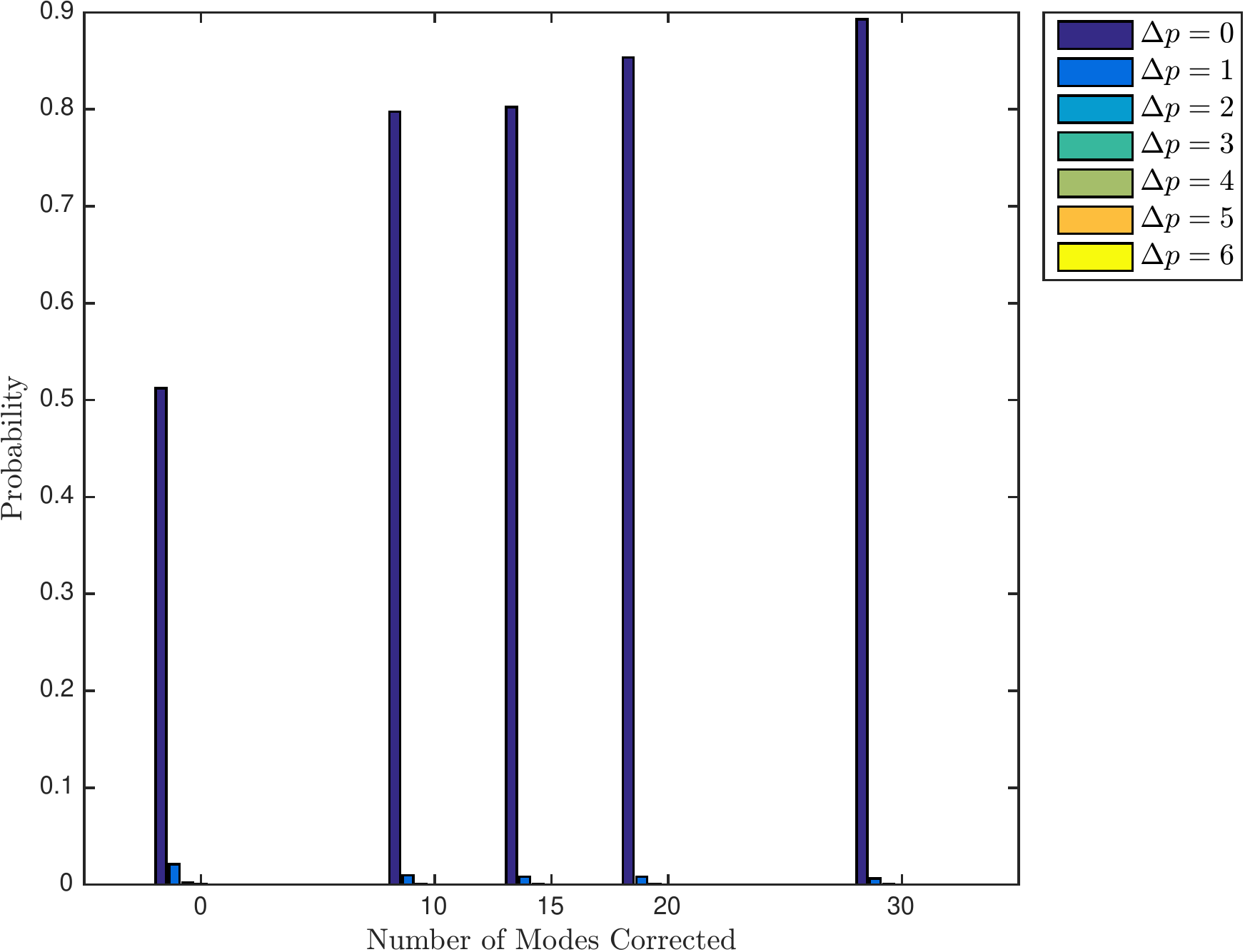}}
\subfigure[$\frac{R}{w(z)}=9.8596$, $\frac{w(z)}{r_0}=0.1167$, and $\frac{z}{z_R}=0.1693$]
{\includegraphics[width=0.45\columnwidth]{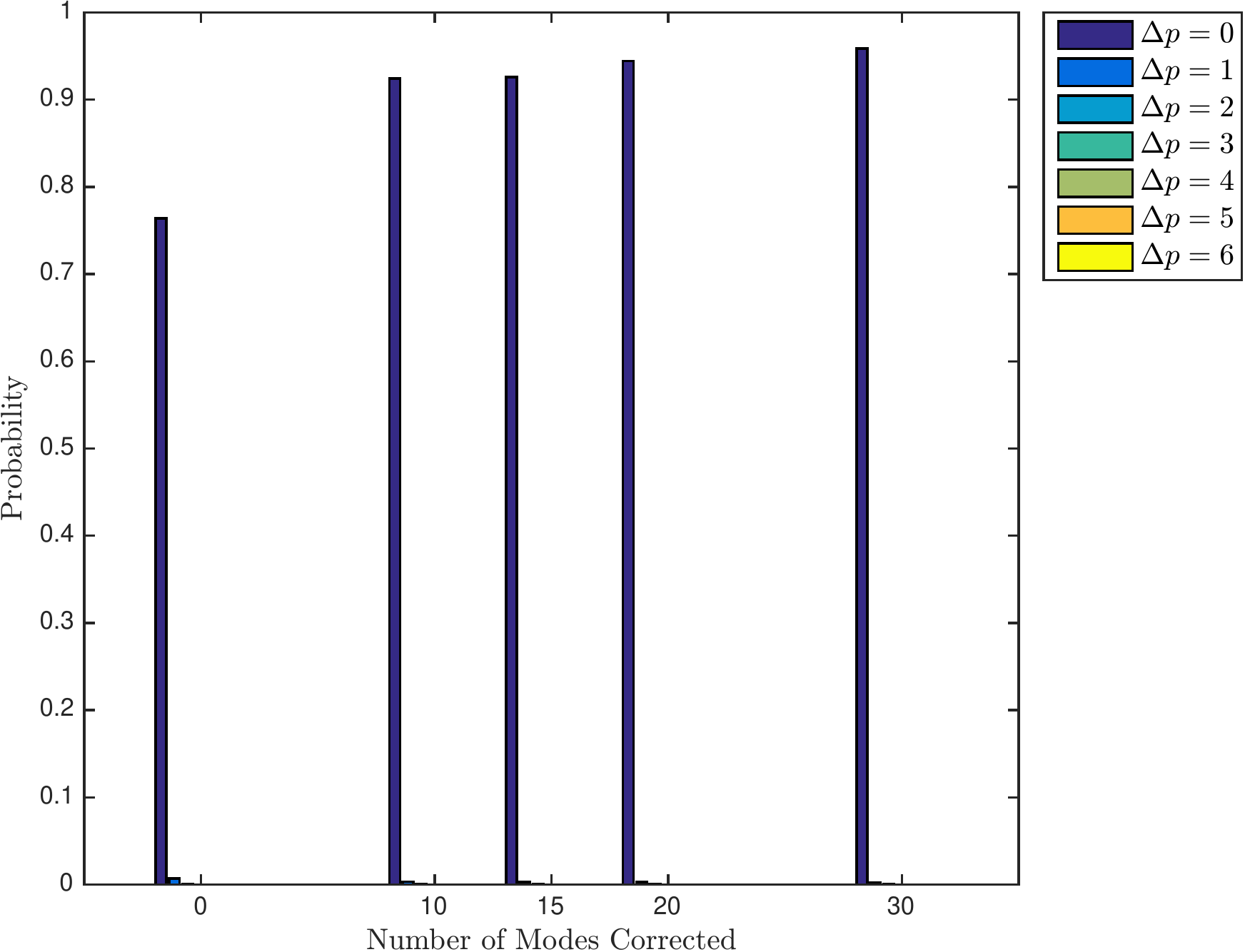}}
\caption{Probability to observe the state $\ket{3,\Delta p}$ after the turbulence and adaptive optics correction for initial state $\ket{3,0}$.}
\label{fig:p-spread}
\end{figure}

We can also study the probability to measure neighboring modes to the initial state after the effects of the channel.  In Figs.~\ref{fig:l-spread} and \ref{fig:p-spread} we see that if the initial state is $\ket{3,0}$, then as we increase the number of modes corrected (or decrease the turbulence strength), the probabilities to observe neighboring states $\ket{3+\Delta l,0}$ or $\ket{3,\Delta p}$ diminish, just as the probability to observe the initial state increases.  Interestingly, the probabilities to observe neighboring azimuthal modes are greater than the probilities to observe neighboring radial modes.

\begin{figure}
\subfigure[$\frac{R}{w(z)}=9.2088$, $\frac{w(z)}{r_0}=0.2165$, and $\frac{z}{z_R}=0.4234$]
{\includegraphics[width=0.45\columnwidth]{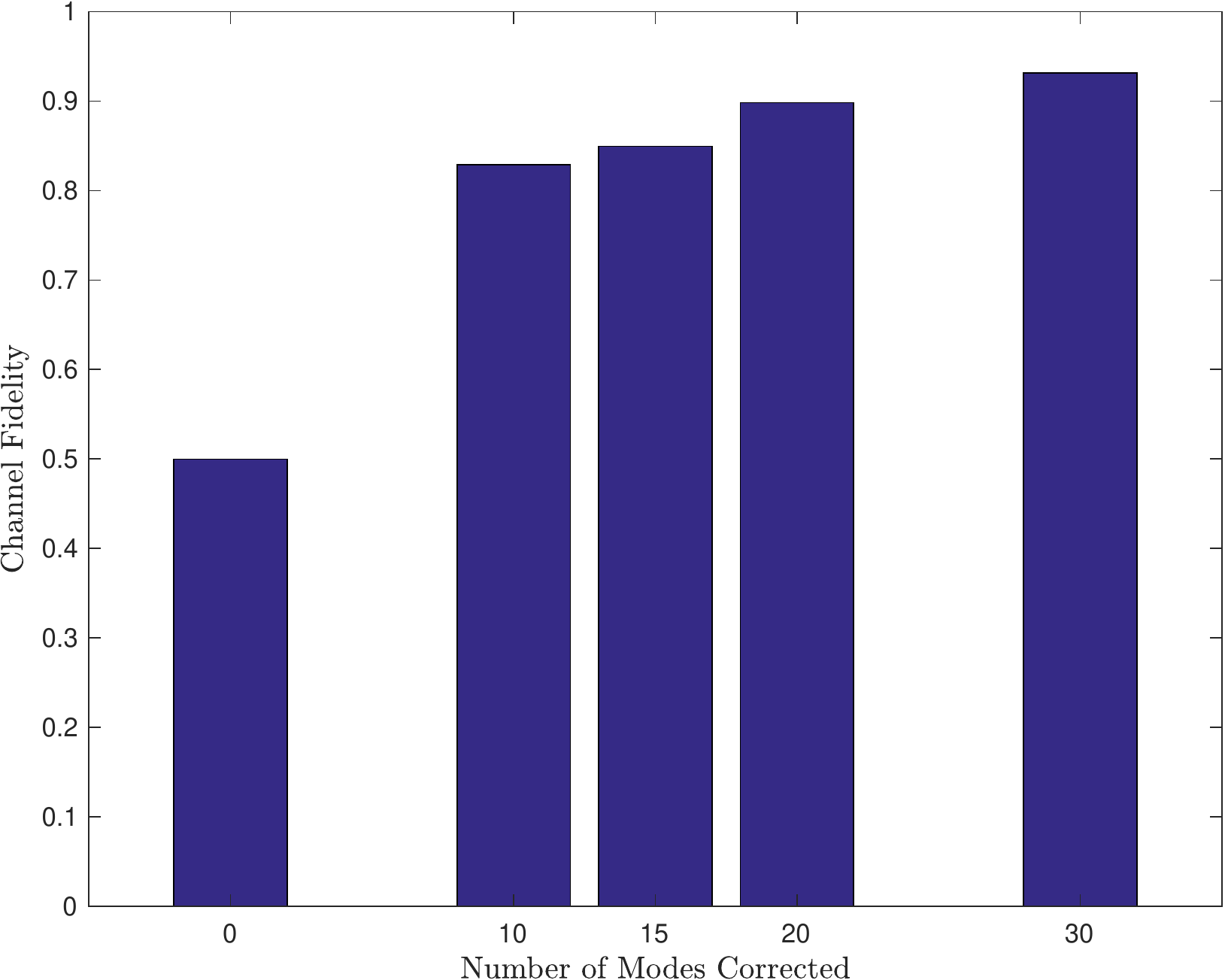}}
\subfigure[$\frac{R}{w(z)}=9.8596$, $\frac{w(z)}{r_0}=0.1167$, and $\frac{z}{z_R}=0.1693$]
{\includegraphics[width=0.45\columnwidth]{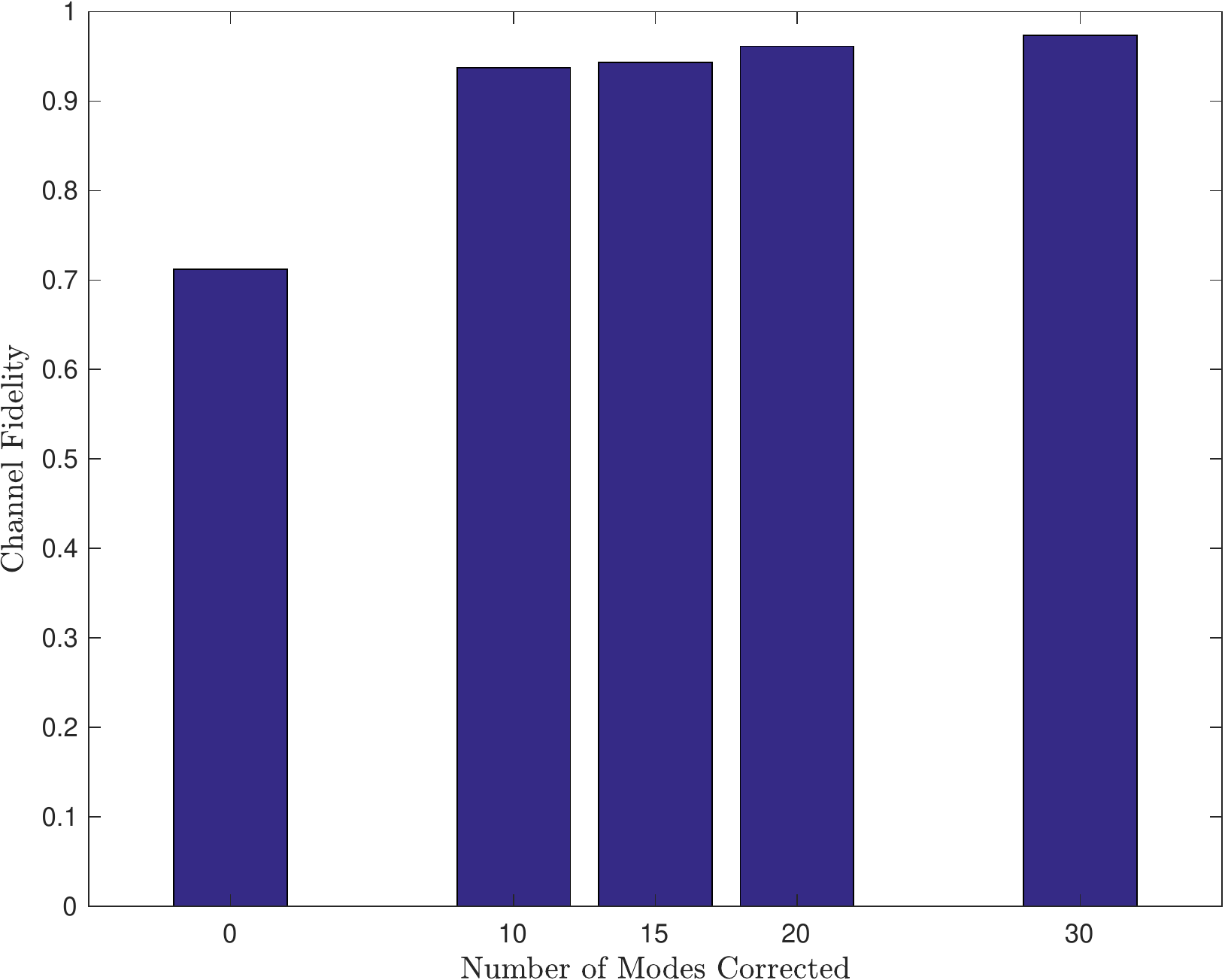}}
\caption{Channel fidelity for the combined effects of turbulence and adaptive optics.}
\label{fig:cf}
\end{figure}

Furthermore, we can also calculate the minimum channel fidelity \cite{Wilde-From-2011-0} for the two turbulence examples we have considered. That is, we
calculate
\begin{align}
F_{\min}(\mathcal{A}_{\ph_A}) = \min_{\ket{\psi}} F\left(\ket{\psi}, \mathcal{A}_{\ph_A}(\ket{\psi}\bra{\psi})\right),
\end{align}
where $F$ is the usual fidelity between quantum states
\begin{align}
F(\rho,\sigma)=\mathrm{Tr}\sqrt{\sqrt{\rho}\sigma\sqrt{\rho}}.
\end{align}
The results for both turbulence examples are shown in Fig. \ref{fig:cf} and Table \ref{tab:cf}. As can be seen, decreasing the 
turbulence strength, or increasing the number of modes corrected, increases the channel fidelity. In particular, we can see that 
increasing the number of modes corrected at first offers a substantial increase in channel fidelity, but later there are diminishing returns with the 
number of modes used. 

\begin{table}
\caption{Channel fidelity results for the combined effects of turbulence and adaptive optics.}
\label{tab:cf}
\begin{center}
\begin{tabular}{ x{0.15\textwidth} | x{0.3\textwidth} | x{0.3\textwidth} }
  Number of modes corrected & Channel fidelity for $\frac{R}{w(z)}=9.2088$, $\frac{w(z)}{r_0}=0.2165$, and $\frac{z}{z_R}=0.4234$ & Channel fidelity for $\frac{R}{w(z)}=9.8596$, $\frac{w(z)}{r_0}=0.1167$, and $\frac{z}{z_R}=0.1693$ \tn
  \hline
   0 & 0.4998 & 0.7119 \tn
  10 & 0.8290 & 0.9374 \tn
  15 & 0.8495 & 0.9432 \tn
  20 & 0.8983 & 0.9614 \tn
  30 & 0.9317 & 0.9738 \tn
\end{tabular}
\end{center}
\end{table}

\section{Conclusions} \label{sec:conclusions}

We have derived an approximate map for the effects of Kolmogorov atmospheric turbulence with adaptive optics on orbital angular momentum states of a photon, assuming that the effects of the uncorrected noise are small.  Using this result, we numerically explored the dependence on some of the dimensionless parameters used to characterize the noise process, and on the number of modes corrected by adaptive optics. We have seen that adaptive optics may compensate for some of the noise due to weak atmospheric turbulence. However, more research is required to extend this to the case of strong turbulence, or to a case where the first order expansion in the covariance is insufficient.

For the case we studied, however, the results seems quite reasonable:  increasing the number of modes corrected by adaptive optics increases the probability that the states are transmitted without errors.  Moreover, it is possible to find a Kraus map for the residual errors after adaptive optics have been applied.  An interesting possibility is to combine adaptive optics (which in a sense reduces the noise in the channel) with quantum error correction (which could be used to protect quantum information from the remaining weak errors).  This is an important direction for future work.

\section{Acknowledgements}
JRGA would like to thank Yongxiong Ren for helpful conversations.  This research was supported by the ARO MURI under Grant No.~W911NF-11-1-0268 and by NSF Grant No.~CCF-1421078.  TAB also acknowledges support as an IBM Einstein Fellow at the Institute for Advanced Study.  Computation for the work in this paper was supported by the University of Southern California Center for High-Performance  Computing and Communications (\url{hpcc.usc.edu}).

\bibliographystyle{apsrev4-1}
\bibliography{../Scientific_References/Reference_Database}

\end{document}